\documentclass[aps,prd,nofootinbib,twocolumn,showkeys,floatfix]{revtex4}
\usepackage{amsmath}
\usepackage{amssymb}
\usepackage{graphicx}
\usepackage{rotating}
\usepackage{color} 
\usepackage{multirow} 
\usepackage[T1]{fontenc} 
\usepackage{slashed}
\newcommand {\black} {\color{black}}

\def\vev#1{\left\langle #1\right\rangle}

\def\321{SU(3) $\otimes$ SU(2) $\otimes$ U(1)}

\newcommand{\AddrAHEP}{
  {\it AHEP Group, Instituto de F\'{\i}sica Corpuscular --
    C.S.I.C./Universitat de Val{\`e}ncia \\
    Edificio de Institutos de Paterna, Apartado 22085,
  E--46071 Val{\`e}ncia, Spain}}
\newcommand{\AddrNiigata}{%
Department of Physics, Niigata University, Niigata 950-2181, Japan.}
\def\gsim{\raise0.3ex\hbox{$\;>$\kern-0.75em\raise-1.1ex\hbox{$\sim\;$}}}
\def\lsim{\raise0.3ex\hbox{$\;<$\kern-0.75em\raise-1.1ex\hbox{$\sim\;$}}}

\begin{document}

\preprint{IFIC/12-23}  

\title{ Predictive Discrete Dark Matter Model}
\author{M.~S. Boucenna} \email{boucenna@ific.uv.es} \affiliation{\AddrAHEP}
\author{S. Morisi} \email{morisi@ific.uv.es} \affiliation{\AddrAHEP}
\author{E. Peinado} \email{eduardo@ific.uv.es} \affiliation{\AddrAHEP}
\author{Yusuke Shimizu} \email{shimizu@muse.sc.niigata-u.ac.jp} \affiliation{\AddrNiigata
}
\author{J.~W.~F.~Valle} \email{valle@ific.uv.es} \affiliation{\AddrAHEP}



\begin{abstract}
 
  Dark Matter stability can be achieved through a partial breaking of
  a flavor symmetry. In this framework we propose a type-II seesaw
  model where left-handed matter transforms nontrivially under the
  flavor group $\Delta(54)$, providing correlations between neutrino
  oscillation parameters, consistent with the recent Daya-Bay and RENO
  reactor angle measurements, as well as lower bounds for neutrinoless
  double beta decay. The dark matter phenomenology is provided by a
  Higgs-portal.

\end{abstract}

\maketitle

The discovery of neutrino oscillations~\cite{art:2012} and the growing
evidence for the existence of dark matter~\cite{Bertone2005279}
provide strong indications for the need of physics beyond Standard
Model (SM). However the detailed nature of the new physics remains
elusive.
On the one hand, the typology of mechanism responsible for neutrino
mass generation and its flavor structure, as well as the nature of the
associated messenger particle are unknown.
Consequently the nature of neutrinos, their mass and mixing parameters
are all unpredicted.

Likewise the nature of Dark Matter (DM) constitutes one of the most
challenging questions in cosmology since decades, though recently some
direct and indirect DM detection experiments are showing tantalizing
hints favoring a light WIMP-like DM candidate \cite{Aalseth:2011wp,Ahmed:2010wy,Angloher:2011uu,Bernabei:2008yi,Hooper:2010mq}
opening hopes for an imminent detection.

Linking neutrino mass generation to dark matter, two seemingly
unrelated problems into a single framework is not only theoretically
more appealing, but also may bring us new insights on both issues.

Among the requirements a viable DM candidate must pass, stability has
traditionally been ensured through the \textit{ad hoc} imposition of a
stabilizing symmetry; usually a parity. Clearly a top-down approach
where stability is naturally achieved is theoretically more appealing.
This is what motivated attempts such gauged as
$U(1)_{B-L}$~\cite{Hambye:2010zb}, gauged discrete
symmetries~\cite{Batell:2010bp} and the recently proposed discrete
dark matter mechanism
(DDM)~\cite{Hirsch:2010ru,Meloni:2011cc,Boucenna:2011tj,Meloni:2010sk},
where stability arises as a remnant of a suitable flavor
symmetry~\footnote{For other flavor models with DM candidates see
  \cite{Meloni:2011cc,Kajiyama:2011gu,Kajiyama:2011fx,Daikoku:2011mq,Kajiyama:2010sb,Haba:2010ag}}.

In its minimal realization, the DDM scenario provides a link between
DM and neutrino phenomenology through the stability issue.  Here we
describe a DDM scenario which is able to connect the two sectors in a
nontrivial way.  The main idea behind DDM is outlined below.

Consider the group of the even permutations of four objects $A_4$.  It
has one triplet and three singlet irreducible representations, denoted
${\bf 3}$ and ${\bf 1,1',1''}$ respectively.  $A_4$ can be broken
spontaneously to one of its $Z_2$ subgroups. Two of the components of
any $A_4$ triplet are odd under such a parity, while the $A_4$ singlet
representation is even. This residual $Z_2$ parity can be used to
stabilize the DM which, in this case, must belong to an $A_4$ triplet
representation, taken as an $SU(2)_L$ scalar Higgs doublet, $\eta\sim
{\bf 3}$
\cite{Hirsch:2010ru,Meloni:2011cc,Boucenna:2011tj,Meloni:2010sk}.
Assuming that the lepton doublets $L_i$ are singlets of $A_4$ while
right-handed neutrinos transform as $A_4$ triplets $N\sim{\bf 3}$, the
contraction rules imply that the DM couples only to Higgses and heavy
right-handed neutrinos $\overline{L}_i \, N\, \tilde{\eta}$.
In this case $\eta$ and $N$ have even as well as odd-components while
$L_i$ are even so that $\overline{L}_i \, N\, \tilde{\eta}$
interaction preserves the $Z_2$ parity.  Invariance under $Z_2$
implies that $N$ components odd under $Z_2$ are not mixed with the
$Z_2$-even light neutrinos $\nu_i$. This forbids the decay of the
lightest $Z_2$-odd component of $\eta$ to light neutrinos through the
heavy right handed neutrinos, ensuring DM stability.
However, simplest schemes of this type lead to $\theta_{13}=0$ as a
first-order prediction~\cite{Hirsch:2010ru}, at variance with recent
reactor results~\cite{An:2012eh,collaboration:2012nd}.

In contrast, assigning the three left-handed leptons to a
flavor-triplet implies that the ``would-be'' DM candidate decays very
fast into light leptons, through the contraction of the triplet
representations, see general discussion in
ref.~\cite{Kajiyama:2011gu}.
This problem has been considered by Eby and Framptom \cite{Eby:2011qa}
using a $T'$ flavor symmetry. While the suggested model has the merit
of incorporating quarks nontrivially, it requires an ``external''
$Z_2$ asymmetry in order to stabilize dark matter.
In fact this observation lead ref.~\cite{Lavoura:2011ry} to claim that
a successful realization of the DDM scenario requires the lepton
doublets to be in three inequivalent singlet representations of the
flavour group.

Here we provide an explicit example of a model based on a $\Delta(54)$
flavour symmetry in which left-handed leptons are assigned to
nontrivial representations of the flavour group, with a viable stable
dark matter particle and a nontrivial inclusion of quarks. In contrast
to the simplest ``flavour-blind'' inert dark matter
scheme~\cite{barbieri:2006dq} our model implies
restrictions and/or correlations amongst the neutrino oscillation
parameters, consistent with the recent reactor angle
measurements~\cite{An:2012eh,collaboration:2012nd}.  Needless to say
the later will pave the way towards a new era for neutrino
oscillations studies~\cite{nunokawa:2007qh,Bandyopadhyay:2007kx}, in
which leptonic CP violation searches will play an important role,
providing an additional motivation for our proposal.


We search for a group $G$ that contains at least two irreducible
representations of dimension larger than one, namely $r_a$ and $r_b$
with dim$(r_{a,b})>1$. We also require that all the components of
$r_a$ transform trivially under an abelian subgroup of $G\supset Z_N$
(with $N=2,3$) while at least one component of $r_b$ is charged with
respect to $Z_N$. The stability of the lightest component of $r_b$ is
guaranteed by $Z_N$ giving a potential~\footnote{Of course,
  other requirements are necessary in order to have a viable DM
  candidate, such as neutrality, correct relic abundance, and
  consistency with constraints from DM search experiments.} DM
candidate.

The simplest group we have found with this feature is $\Delta(54)$,
isomorphic to $(Z_3\times Z_3) \rtimes S_3$.  In addition to the
irreducible triplet representations, $\Delta(54)$ contains four
different doublets ${\bf 2_{1,2,3,4}}$ and two irreducible singlet
representations, ${\bf 1_{\pm}}$.  The product rules for the doublets
are ${\bf 2_k}\times {\bf 2_k}= {\bf 1_+}+{\bf 1_-}+{\bf 2_k}$ and
${\bf 2_1}\times {\bf 2_2}={\bf 2_3}+{\bf 2_4}$. Of the four doublets
${\bf 2_1}$ is invariant under the $P\equiv (Z_3\times Z_3)$ subgroup
of $\Delta(54)$, while the others transform nontrivially, for example
${\bf 2_3} \sim (\chi_1,\chi_2)$, which transforms as
$\chi_1\,(\omega^2,\omega)$ and $\chi_2\,(\omega,\omega^2)$
respectively, where $\omega^3=1$
\cite{Ishimori:2010au,Ishimori:2008uc}. We can see that by taking $r_a
= {\bf 2_1}$ and $r_b = {\bf 2_3}$ that $\Delta(54)$ is perfect for
our purpose.


Let us now turn to the explicit model, described in table~\ref{tab1},
where $L_D \equiv (L_\mu ,L_\tau) $ and $l_D \equiv (\mu_R, \tau_R)$.
There are 5 $SU(2)_L$ doublets of Higgs scalars: the $H$ is a singlet
of $\Delta(54)$, while $\eta=(\eta_1,\eta_2) \sim {\bf 2_3}$ and
$\chi=(\chi_1,\chi_2) \sim {\bf 2_1}$ are doublets.  In order to
preserve a remnant $P$ symmetry, the doublet $\eta$ is not allowed to
take vacuum expectation value (vev).
Such a prescription is not necessary for $H$, $\chi_1$ and $\chi_2$
since these are all invariant under $P$.  We also need to introduce an
$SU_L(2)$ Higgs triplet scalar field $\Delta \sim{\bf 2_1}$ whose vev
will induce neutrino masses through the type-II seesaw
mechanism~\cite{schechter:1980gr}.
Regarding dark matter, note that the lightest $P$-charged particle in
$\eta_{1,2}$ can play the role of ``inert'' DM~\cite{barbieri:2006dq},
as it has no direct couplings to matter. The conceptual link
between dark matter and neutrino phenomenology arises from the fact
that the DM stabilizing symmetry is a remnant of the underlying flavor
symmetry which accounts for the observed pattern of oscillations.  See
phenomenological implications below.
\begin{table}[h]
\begin{center}
\begin{tabular}{ccccc|cccc}
\hline
\hline
& $\overline{L}_e$ & $\overline{L}_D$ & $e_R$ & $l_{D}$   
& $H$  & $\chi $ & $\eta $ &$\Delta$\\
\hline 
$SU(2)$ & $2$ & $2$ & $1$ & $1$  &$2$ & $2$ & $2$ & $3$\\
$\Delta (54)$ & $\bf 1_+$ & $\bf 2_1$ & $\bf 1_+$ & $\bf 2_1$    & $\bf 1_+$& $\bf 2_1$ & $\bf 2_3$ & $\bf 2_1$ \\
\hline
\hline
\end{tabular}
\caption{Lepton and higgs boson assignments of the model. }\label{tab1}
\end{center}
\end{table}

The lepton part of the Yukawa Lagrangian is given by  
\begin{eqnarray}
\mathcal{L}_{\ell}&=&y_1 \overline{L}_e e_R H+
y_2 \overline{L}_e l_D \chi+ 
y_3 \overline{L}_D e_R\,\chi + \\
&&+ y_4\overline{L}_Dl_D H + y_5 \overline{L}_Dl_D \chi \nonumber\\
&&\nonumber\\
\mathcal{L}_{\nu}&=&
y_b\overline{L}_D \overline{L}_D\Delta +
y_a\overline{L}_DL_e \Delta 
\end{eqnarray}
After electroweak symmetry breaking the first term
$\mathcal{L}_{\ell}$ gives the following charged lepton mass matrix
\begin{equation}\label{Ml}
M_\ell =
\begin{pmatrix}
a & br & b \\
cr & d & e \\
c & e & dr
\end{pmatrix}
\end{equation}
where $a=y_1 \vev{H}$, $b= y_2\vev{\chi_1}$, $c=y_3 \vev{\chi_1}$,
$d=y_5 \vev{\chi_1}$, $e=y_4 \vev{H}$, and
$$r=\vev{\chi_2}/\vev{\chi_1}.$$

On the other hand the $\mathcal{L}_{\nu}$ is the term responsible for
generating the neutrino mass matrix.  Choosing the solution
$\vev{\Delta}\sim (1,1)$ and $\vev{\chi_1} \ne \vev{\chi_2}$,
consistent with the minimization of the scalar potential one finds
that
\begin{equation}\label{mnu}
M_\nu \propto 
\begin{pmatrix}
0 & \delta  & \delta   \\
\delta & \alpha & 0 \\
\delta  & 0  & \alpha
\end{pmatrix},
\end{equation}
where $\delta=y_a\vev{\Delta}$, $\alpha= y_b \vev{\Delta}$. 
%

Our model corresponds to a flavour-restricted realization of the {\it
  inert dark matter} scenario proposed in~\cite{barbieri:2006dq}. As
such, it has nontrivial consequences for neutrino phenomenology, which
we now study in detail.
As seen in eq.~(\ref{mnu}) the neutrino mass matrix depends only on
two parameters, $\delta$ and $\alpha$ (taken to be real), which can be
expressed as a function of the measured squared mass differences, as
follows
\begin{equation}
m_{1,3}^\nu=\frac{\alpha \mp \sqrt{8 \delta^2+\alpha^2}}{2}, \: m_2^\nu=\alpha.
\label{massesin}
\end{equation}

\begin{figure}[h!]
\begin{center}
 \includegraphics[width=6.5cm]{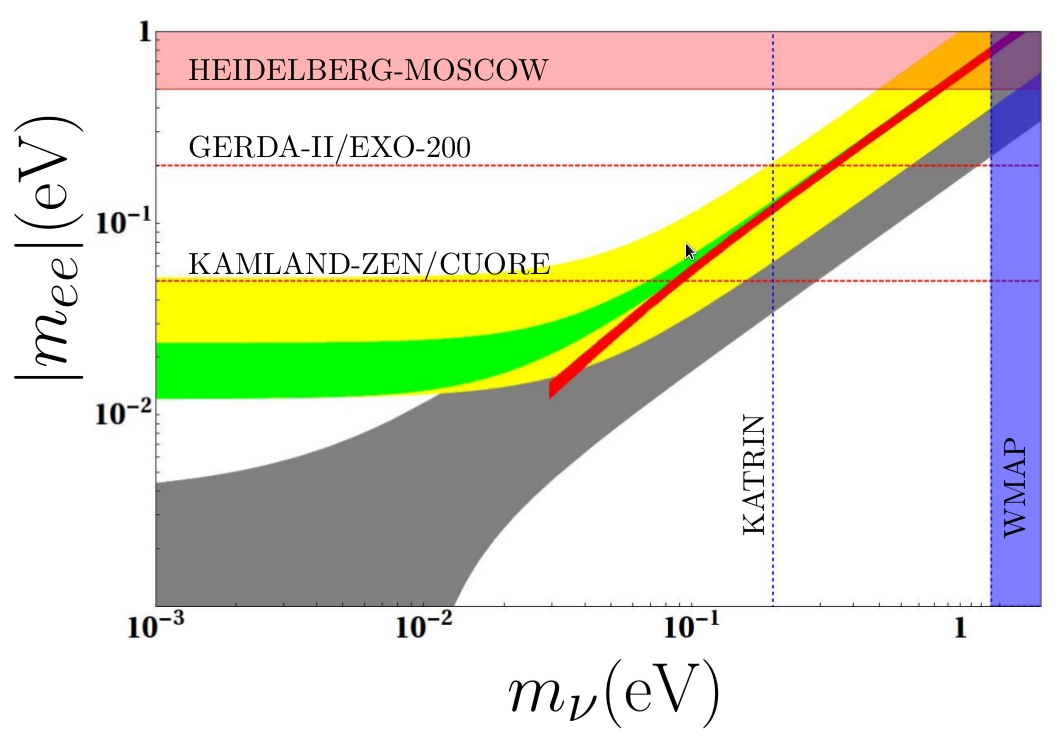}
 \caption{Effective neutrinoless double beta decay parameter $m_{ee}$
   versus the lightest neutrino mass. The thick upper and lower
   branches correspond the ``flavor-generic'' inverse (yellow) and
   normal (gray) hierarchy neutrino spectra, respectively.  The model
   predictions are indicated by the green and red (darker-shaded)
   regions, respectively. Only these sub-bands are allowed by the
   $\Delta (54)$ model.  For comparison we give the current limit and
   future sensitivities on
   $m_{ee}$~\cite{Schwingenheuer:2012jt,Rodejohann:2011mu} and
   $m_\nu$~\cite{Osipowicz:2001sq,Komatsu:2010fb}, respectively.
 }\label{figbb}
\end{center}
\end{figure}

For simplicity, we fix the intrinsic neutrino
CP--signs~\cite{Schechter:1981hw} as $\eta=diag(-,+,+)$, where $\eta$
is defined as $U^\star=U \eta$, $U$ being the lepton mixing matrix.
It is easy to check that, in this case, only a normal hierarchy
spectrum is allowed.
In contrast, a different permutation of the eigenvalues corresponding
to our $\eta$ matrix, namely $(1,2,3) \to (1,3,2)$ in
Eq.\,\ref{massesin}, gives only inverse hierarchy spectrum.
Moreover, notice that the masses in eq.~(\ref{massesin}) obey a
neutrino mass sum rule of the form $m_1^\nu+m_2^\nu=m_3^\nu$ which has
implications for the neutrinoless double beta decay
process~\cite{Dorame:2011eb}, as illustrated in Fig.~(\ref{figbb}).

\begin{figure}[h!]
\begin{center}
 \includegraphics[width=6cm]{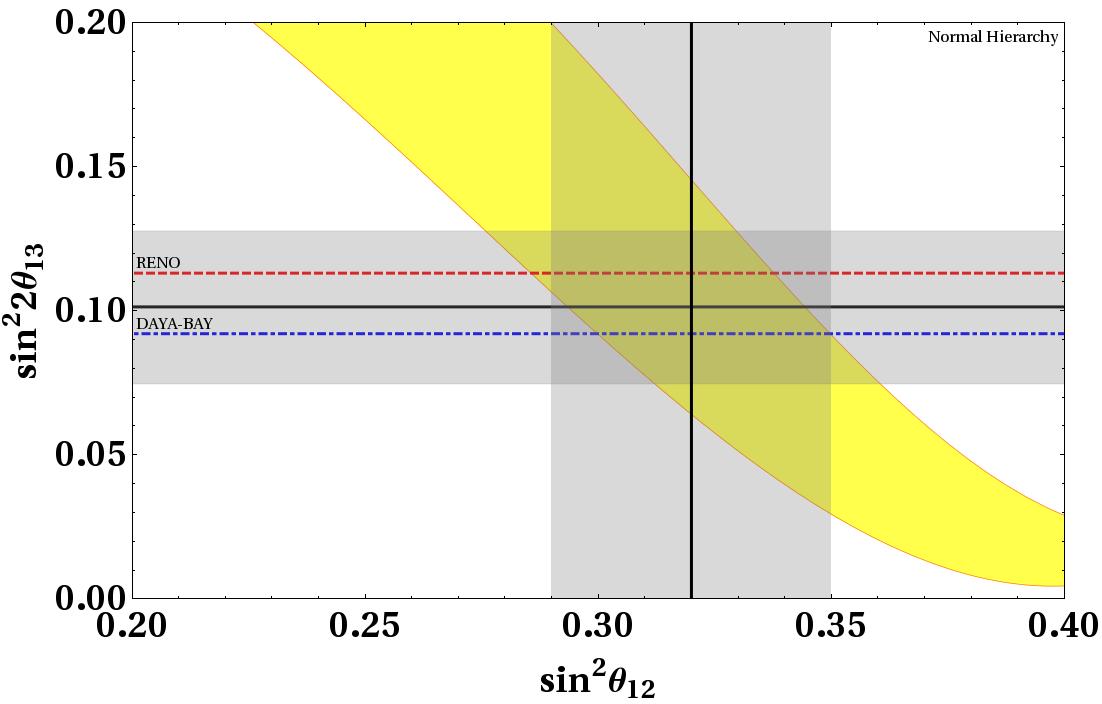}
 \caption{The shaded (yellow) curved band gives the predicted
   correlation between solar and reactor angles when $\theta_{23}$ is
   varied within 2$\sigma$ for the normal hierarchy spectrum. The
   solid (black) line gives the global best fit values for
   $\theta_{12}$ and $\theta_{13}$, along with the corresponding
   two-sigma bands, from Ref.~\cite{Schwetz:2011zk}. The dashed lines
   correspond to the central values of the recent reactor
   measurements~\cite{An:2012eh,collaboration:2012nd}.}\label{figcn}
\end{center}
\end{figure}

\begin{figure}[h!]
\begin{center}
 \includegraphics[width=6cm]{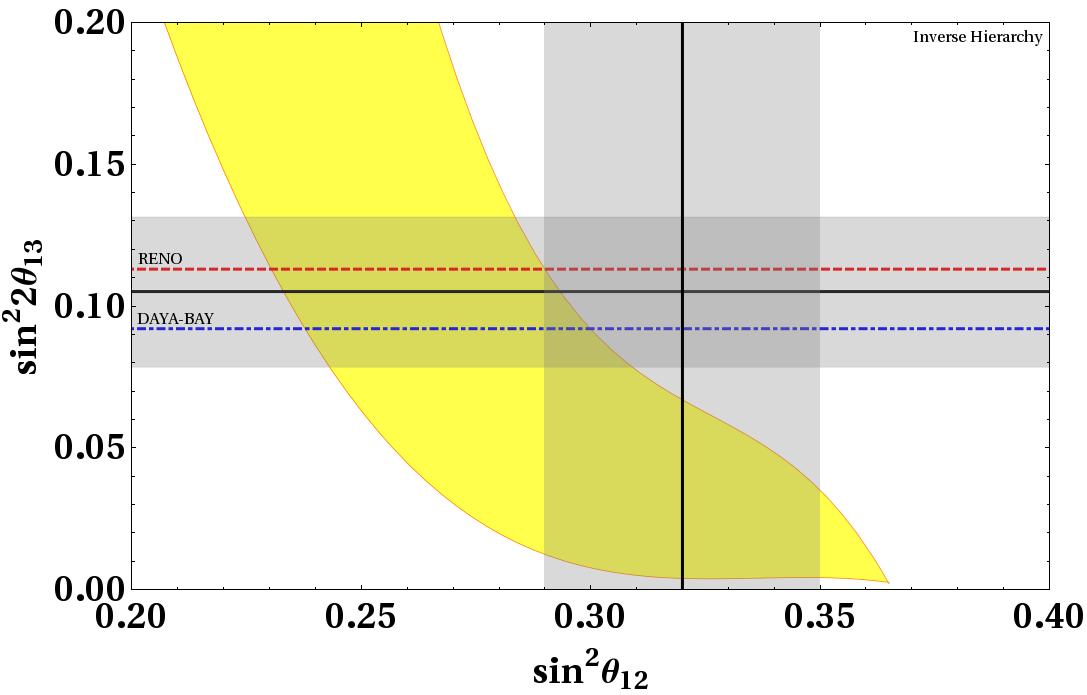}
 \caption{Same as above for the inverse hierarchy case.}\label{figci}
\end{center}
\end{figure}

We now turn to the second prediction. Although in our scheme
neutrino mixing parameters in the lepton mixing matrix are not
strictly predicted, there are correlations between the
reactor and the atmospheric angle, as illustrated in
Figs.~\ref{figcn}~\footnote{There is also a second band allowed in
  this case which is, however, experimentally ruled out by the
  measurents of $\theta_{12}$ and $\theta_{13}$.} and Fig.~\ref{figci}
for the cases of normal and inverse mass hierarchies, respectively.
While the solar angle is clearly unconstrained and can take all the
values within in the experimental limits, correlations
exist with the reactor mixing angle, indicated by the curved yellow
bands in Fig.~\ref{figcn} and \ref{figci}. These correspond to
$2\sigma$ regions of $\theta_{23}$ as determined in
Ref.~\cite{Schwetz:2011zk}.  The horizontal lines give the best global
fit value and the recent best fit values obtained in Daya--Bay and
RENO reactors~\cite{An:2012eh,collaboration:2012nd} (see also recent
result from T2K \cite{Abe:2011sj}).
\black
Now we turn to quarks. In
Ref.~\cite{Hirsch:2010ru,Meloni:2011cc,Boucenna:2011tj,Meloni:2010sk}
quarks were singlets of the flavor symmetry to guarantee the stability
of the DM. Consequently the generation of quark mixing was difficult
\cite{deAdelhartToorop:2011ad}.  This problem has been recently
considered in \cite{Eby:2011qa} using $T'$ flavor symmetry.

A nice feature of our current model is that with $\Delta(54)$ we can
assign quarks to the singlet and doublet representations as shown in
table\,\ref{tab2}. This opens new possibilities to fit the CKM mixing
parameters.
Indeed, as shown in table~\ref{tab2} quarks transforming nontrivially
under the flavor symmetry can be consistently added in our picture.
\begin{table}[h!]
\begin{center}
\begin{tabular}{cccccccc}
\hline
\hline
$ $ & $Q_{1,2}$ & $ Q_{3}$ & $(u_{R},c_R)$ &$t_R$ &$d_R$ & $s_R$ &$b_R$\\
\hline
$SU(2)$ & $2$ & $2$ & $1$ & $1$  &$1$ & $1$ & $1$ \\
\hline
$\Delta(54)$ & ${\bf 2_1}$& ${\bf 1_+}$& ${\bf 2_1}$& ${\bf 1_+}$& ${\bf 1_-}$& ${\bf 1_+}$& ${\bf 1_+}$\\
\hline
\hline
\end{tabular}\caption{Quark gauge and flavour representation assignments.}
\label{tab2}
\end{center}
\end{table}

The resulting up- and down-type quark mass matrices in our model are
given by
\begin{equation}
M_d=
\begin{pmatrix}
ra_d & rb_d & rd_d \\
-a_d & b_d & d_d \\
0 & c_d & e_d
\end{pmatrix},\quad 
M_u=
\begin{pmatrix}
ra_u & b_u & d_u \\
b_u & a_u & rd_u \\
c_u & rc_u & e_u
\end{pmatrix}.
\end{equation}

Note that the Higgs fields $H$ and $\chi$ are common to the lepton and
the quark sectors and in particular the parameter $r$.  
Assuming for simplicity real couplings we have 11 free parameters
characterizing this sector, 10 Yukawa couplings plus the ratio of the
the isodoublet vevs, $r$, introduced earlier in the neutrino
sector.
We have verified that we can make a fit of all quark masses and
mixings provided $r$ lies in the range of about $ 0.1<r<0.2$. We do
not extend further the discussion on the quark interactions which can
be easily obtained from table\,\ref{tab2} (a full analysis of the
quark phenomenology is beyond the scope of this paper and will be
taken up elsewhere).


Notice that our scalar Dark matter candidate $\eta_1$ has quartic
couplings with the Higgs scalars of the model such as
$\eta^\dagger\eta \,H^\dagger H$ and $\eta^\dagger\eta \,\chi^\dagger
\chi$. These weak strength couplings provide a Higgs portal production
mechanism, and ensure an adequate cosmological relic abundance.
Direct and indirect detection prospects are similar to a generic WIMP
dark matter, as provided by multi-Higgs extensions of the SM.

In short we have described how spontaneous breaking of a $\Delta(54)$
flavor symmetry can stabilize the dark matter by means of a residual
unbroken symmetry. In our scheme left-handed leptons as well as quarks
transform nontrivially under the flavor group, with neutrino masses
arising from a type-II seesaw mechanism.  We have found 
lower bounds for neutrinoless double beta decay, even in the case of
normal hierarchy, as seen in Fig.~\ref{figbb}. In addition, we have
correlations between solar and reactor angles consistent with the
recent Daya-Bay and RENO reactor measurements, see Fig.~\ref{figcn}
and Fig.~\ref{figci}.

\black
Unfortunately, however, the DM particle is not directly involved as
messenger in the neutrino mass generation mechanism. This issue will
be considered elsewhere.

\black

We thank Luis Dorame, Luis Lavoura and Martin Hirsch for discussions
and interesting comments. This work was supported by the Spanish MEC
under grants FPA2011-22975 and MULTIDARK CSD2009-00064
(Consolider-Ingenio 2010 Programme), by Prometeo/2009/091 (Generalitat
Valenciana), by the EU ITN UNILHC PITN-GA-2009-237920. Y.S. is
supported by Grand-in-Aid for Scientific Research, No.22.3014 in JSPS.

\bibliographystyle{h-physrev4}

\begin{thebibliography}{10}

\bibitem{art:2012} A.~McDonald, \newblock Talk at XIV International
  Workshop on Neutrino Telescopes, Venice, March, 2011; for a review
  see M.~Maltoni et al, New J.\ Phys.\ {\bf 6} (2004) 122.

\bibitem{Bertone2005279}
G.~Bertone, D.~Hooper and J.~Silk,
\newblock Physics Reports {\bf 405}, 279  (2005).

\bibitem{Aalseth:2011wp}
C.~Aalseth {\em et~al.},
\newblock Phys.Rev.Lett. {\bf 107}, 141301 (2011), [1106.0650].

\bibitem{Bernabei:2008yi}
  R.~Bernabei {\it et al.}  [DAMA Collaboration],
  Eur.\ Phys.\ J.\ C {\bf 56} (2008) 333
  [arXiv:0804.2741 [astro-ph]].
\bibitem{Angloher:2011uu}
  G.~Angloher, M.~Bauer, I.~Bavykina, A.~Bento, C.~Bucci, C.~Ciemniak, G.~Deuter and F.~von Feilitzsch {\it et al.},
  arXiv:1109.0702 [astro-ph.CO].

\bibitem{Ahmed:2010wy}
  Z.~Ahmed {\it et al.}  [CDMS-II Collaboration],
  Phys.\ Rev.\ Lett.\  {\bf 106} (2011) 131302
  [arXiv:1011.2482 [astro-ph.CO]].

\bibitem{Hooper:2010mq}
  D.~Hooper and L.~Goodenough,
  Phys.\ Lett.\ B {\bf 697} (2011) 412
  [arXiv:1010.2752 [hep-ph]].


\bibitem{Hambye:2010zb}
T.~Hambye,
\newblock PoS {\bf IDM2010}, 098 (2011), [1012.4587].

\bibitem{Batell:2010bp}
B.~Batell,
\newblock Phys.Rev. {\bf D83}, 035006 (2011), [1007.0045].

\bibitem{Hirsch:2010ru}
M.~Hirsch, S.~Morisi, E.~Peinado and J.~W.~F.~Valle,
\newblock Phys.Rev. {\bf D82}, 116003 (2010), [1007.0871].

\bibitem{Meloni:2011cc}
D.~Meloni, S.~Morisi and E.~Peinado,
\newblock Phys.Lett. {\bf B703}, 281 (2011), [1104.0178].

\bibitem{Boucenna:2011tj}
M.~Boucenna {\em et~al.},
\newblock JHEP {\bf 1105}, 037 (2011), [1101.2874].

\bibitem{Meloni:2010sk}
D.~Meloni, S.~Morisi and E.~Peinado,
\newblock Phys.Lett. {\bf B697}, 339 (2011), [1011.1371].

\bibitem{Kajiyama:2011gu}
Y.~Kajiyama, K.~Kannike and M.~Raidal,
\newblock Phys.Rev. {\bf D85}, 033008 (2012), [1111.1270],
\newblock 8 pages, no figures.

\bibitem{Kajiyama:2011fx}
Y.~Kajiyama, H.~Okada and T.~Toma,
\newblock 1109.2722.

\bibitem{Daikoku:2011mq}
Y.~Daikoku, H.~Okada and T.~Toma,
\newblock Prog.Theor.Phys. {\bf 126}, 855 (2011), [1106.4717].

\bibitem{Kajiyama:2010sb}
Y.~Kajiyama and H.~Okada,
\newblock Nucl.Phys. {\bf B848}, 303 (2011), [1011.5753].

\bibitem{Haba:2010ag}
N.~Haba, Y.~Kajiyama, S.~Matsumoto, H.~Okada and K.~Yoshioka,
\newblock Phys.Lett. {\bf B695}, 476 (2011), [1008.4777].

\bibitem{An:2012eh}
Daya-Bay Collaboration, F.~An {\em et~al.},
\newblock 1203.1669,
\newblock 5 figures.

\bibitem{collaboration:2012nd}
Soo-Bong Kim for RENO collaboration, 
\newblock 1204.0626.

\bibitem{Eby:2011qa}
D.~A. Eby and P.~H. Frampton,
\newblock 1111.4938.

\bibitem{Lavoura:2011ry}
L.~Lavoura,
\newblock J.Phys.G {\bf G39}, 025202 (2012), [1109.6854].

\bibitem{barbieri:2006dq}
R.~Barbieri, L.~J. Hall and V.~S. Rychkov,
\newblock Phys. Rev. {\bf D74}, 015007 (2006), [hep-ph/0603188].

\bibitem{Schechter:1981hw}
  J.~Schechter and J.~W.~F.~Valle,
\newblock  Phys.\ Rev.\ D {\bf 24} (1981) 1883
   [Erratum-ibid.\ D {\bf 25} (1982) 283].

\bibitem{nunokawa:2007qh}
H.~Nunokawa, S.~J. Parke and J.~W.~F. Valle,
\newblock Prog. Part. Nucl. Phys. {\bf 60}, 338 (2008).

\bibitem{Bandyopadhyay:2007kx}
ISS Physics Working Group, A.~Bandyopadhyay {\em et~al.},
\newblock Rept.Prog.Phys. {\bf 72}, 106201 (2009), [0710.4947].

\bibitem{Ishimori:2010au}
H.~Ishimori {\em et~al.},
\newblock Prog. Theor. Phys. Suppl. {\bf 183}, 1 (2010), [1003.3552].

\bibitem{Ishimori:2008uc}
H.~Ishimori, T.~Kobayashi, H.~Okada, Y.~Shimizu and M.~Tanimoto,
\newblock JHEP {\bf 0904}, 011 (2009), [0811.4683].

\bibitem{schechter:1980gr}
J.~Schechter and J.~W.~F. Valle,
\newblock Phys. Rev. {\bf D22}, 2227 (1980);
\newblock Phys. Rev. {\bf D25}, 774 (1982).

\bibitem{Schwetz:2011zk} M.~Tortola, J.~W.~F. Valle, and D. Vanegas,
  arXiv:1205.4018; which updates previous analysis before the recent
  reactor results, e.~g. T.~Schwetz, M.~Tortola and J.~W.~F. Valle,
  \newblock New J. Phys. {\bf 13}, 063004 (2011);
\newblock New J.Phys. {\bf 13}, 109401 (2011).

\bibitem{Dorame:2011eb}
L.~Dorame, D.~Meloni, S.~Morisi, E.~Peinado and J.~Valle,
\newblock 1111.5614.

\bibitem{Schwingenheuer:2012jt} 
  B.~Schwingenheuer,
  arXiv:1201.4916 [hep-ex].

\bibitem{Rodejohann:2011mu}
  W.~Rodejohann,
  Int.\ J.\ Mod.\ Phys.\ E {\bf 20} (2011) 1833
  [arXiv:1106.1334 [hep-ph]].

\bibitem{Komatsu:2010fb}
  E.~Komatsu {\it et al.}  [WMAP Collaboration],
  Astrophys.\ J.\ Suppl.\  {\bf 192} (2011) 18
  [arXiv:1001.4538 [astro-ph.CO]].

\bibitem{Osipowicz:2001sq}
  A.~Osipowicz {\it et al.}  [KATRIN Collaboration],
  hep-ex/0109033.

\bibitem{Abe:2011sj}
T2K Collaboration, K.~Abe {\em et~al.},
\newblock Phys.Rev.Lett. {\bf 107}, 041801 (2011), [1106.2822].

\bibitem{deAdelhartToorop:2011ad}
R.~de~Adelhart~Toorop, F.~Bazzocchi and S.~Morisi,
\newblock Nucl.Phys. {\bf B856}, 670 (2012), [1104.5676].

\end{thebibliography}

\end{document}